# NN Based Active Disturbance Rejection Controller for a Multi-Axis Gimbal System

Damla Leblebicioglu[1, 2], Ozgur Atesoglu[1, 2] and Melih Cakmakci[2]

*Abstract*— The increasing demand for target tracking, environmental surveys, surveillance and mapping requires multi-axis gimbal systems with high tracking and stabilization performance. In this paper, first, *computed torque model* is generated to estimate the complex disturbances acting on the system. Then, two different control strategies based on active disturbance rejection control (ADRC) and *computed torque model* are implemented on a two-axis gimbal system. The purpose is to improve the robustness, environmental adaptability and tracking accuracy of the system and reduce the tuning effort of ADRC by integrating a neural network (NN) based disturbance compensator (NN assisted ADRC). In the second control strategy, NN is replaced with a *computed torque model* (CTM assisted ADRC), whose inputs come from plant outputs. The simulation results show that, NN and CTM assisted ADRC structures can decrease mean tracking errors up to 85.4% and 40.8%, respectively.

## I. INTRODUCTION

Multi-axis gimbal systems are frequently used as mounts for day and infrared cameras [1, 2]. This study concentrates on the modeling and proposes an alternative control methodology of a multi-axis gimbal system that will be mounted on an unmanned aerial vehicle [3].

### A. Motivation

The objective is to improve the position control loop performance of the gimbal system that is currently working with conventional ADRC by adding a coupled, neural network based disturbance compensator. NN supports the ADRC by learning the acceleration required to compensate the loss arising from external, internal disturbances and parameter uncertainties of the system. As a result, gimbal follows the reference position commands ensuring higher precision on reference tracking and inner loop stabilization.

### B. Related Work

A two-axis gimbal is a complex, multiple-input and multiple-output (MIMO), nonlinear system that is susceptible to effects of unknown frictional torques, external/internal disturbances and modeling uncertainties. Various control strategies [4, 5, 6, 7, 8, 9] have been developed to satisfy the accurate positioning requirements during stabilization and tracking related missions. Among these, a popular and a powerful technique is ADRC, which first estimates disturbances using ESO (extended state observer). It is proposed by J. Han in [10]. ADRC is a controller with sufficient adaptability that improves dynamic performance of the system. It is a method proven to be a suitable alternative for the conventional proportional-integral-derivative (PID) controller. PID controllers are widely applicable to various systems due to their straightforward structure and comparably effortless design. Nonetheless, they are insufficient for compensation when nonlinear disturbances and parameter uncertainties are significant. ADRC inherits the best parts and eliminates the limitations of a simple PID controller. Moreover, this method needs very little model information [11, 12, 13, 14]. However, ADRC necessitates a certain parameter set to be adjusted for the best design, which is a challenging and a time-consuming task [13, 14]. Traditional ADRC structure is quite sensitive to changes in model parameters, states of the system and inputs. Besides, controller parameters have a strong influence on the robustness and disturbance rejection capability of the system, [14, 15, 16, 17, 18]. In recent years, many researchers produced quite successful back-up solutions for these problems, especially by using NNs. For example, a classical state observer and a radial basis function NN (RBFNN) is used to estimate the unknown total disturbance [15]. Similarly, an RBFNN is used in a dual channel composite and adaptive controller scheme without a state observer in [16]. Tuning problem is solved by using a NN which outputs ADRC parameters in response to changes in plant parameters and disturbances [17].

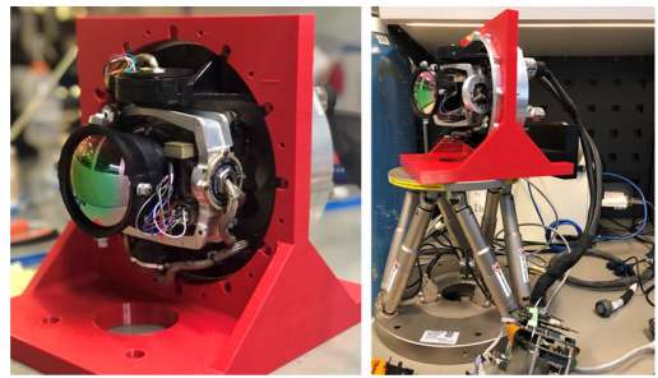

Fig. 1. Two-axis gimbal system

[1] D. Leblebicioglu, O. Atesoglu, A.E. Derinoz are with the Roketsan Missiles Inc., 06780 Ankara, Turkey, {damla.leblebicioglu, ozgur.atesoglu, anil.derinoz}@roketsan.com.tr.

[2] D. Leblebicioglu, O. Atesoglu, M. Cakmakci are with the Department of Mechanical Engineering, Bilkent University, 06800 Ankara, Turkey, {d.leblebicioglu, oatesoglu, melihc}@bilkent.edu.tr.

## C. Original Contributions

In this study, first an accurate gimbal model is implemented in Simulink® and verified with the hardware set-up in Fig. 1. A standard ADRC structure in [10] is implemented by using the *Computed Torque Model (CTM)*. Afterwards, a NN whose inputs are plant position, velocity and acceleration and whose output is the delta (difference between CTM and real plant) acceleration commands to support the ADRC output has been embodied to ADRC. This new structure: (i) made parameter tuning of ADRC much easier, (ii) improves tracking accuracy, (iii) increases disturbance rejection ability of the system. It works as a complimentary back-up when ADRC is not completely sufficient to meet the desired performance requirements. Also, a possible real-time tuning and adjustment of NN provides enhanced adaptivity of the control system under internal/external and time varying changes on the system (iv).

In addition, ADRC controller uses distinct ESOs for each channel. Instead, there is only one NN and it works as a MIMO disturbance compensator. Inputs of the NN are fed from both the azimuth and elevation axes, and outputs of the NN are supplementary accelerations for both channels. Due to authors' knowledge, the proposed controller and compensation scheme is applied on a two-axis gimbal system for the first time. Furthermore, it is relatively simpler when compared to existing similar studies, [15, 16, 18]. Moreover, another approach is also presented as an alternative to NN: necessary compensation can be supplied by adding a second *CTM block* for the ADRC structure (*CTM* based ADRC controller) to replace the NN.

## D. Organization

The paper is organized as follows. Section II introduces the mathematical model of the gimbal dynamics and the *Computed Torque Model* for a two-axis gimbal system. Section III presents the classical ADRC, NN assisted ADRC and *CTM* assisted ADRC designs for the multi-axis gimbal system model. Section IV summarizes and compares simulation results for the suggested controllers. The final section (Section V) lists conclusions and future work about the current implementation.

## II. MATHEMATICAL MODEL OF THE SYSTEM

The mechanical structure of the two-axis gimbal system in Fig. 1 is composed of three interconnected rigid bodies: the inner ring (for the elevation motion), the outer ring (for the azimuth motion) and the base platform. The sighting camera and the micro-electromechanical system (MEMS) gyroscope are installed on the inner ring. Thus, they provide measurements corresponding to the inner gimbal reference frame $F_m$ (with axes $x_m, y_m, z_m$). Outer gimbal rotates with angle $\psi_a$ around the $z_b$-axis of the base platform reference frame $F_b$ (with axes $x_b, y_b, z_b$). Inner gimbal rotates with angle $\theta_m$ around the $y_a$-axis of the outer gimbal reference frame $F_a$ (with axes $x_a, y_a, z_a$). The Euler angles of the base platform with respect to (wrt.) a chosen inertial reference frame are denoted by $\phi, \theta, \psi$. The angular velocity ($\bar{\omega}_{b/o}^{(b)}$) and translational acceleration ($\bar{a}_{b/o}^{(b)}$) of the base platform wrt. a chosen inertial reference frame, expressed in the base platform reference frame are $\bar{\omega}_{b/o}^{(b)} = [p\ q\ r]^T$ and $\bar{a}_{b/o}^{(b)} = [a_x\ a_y\ a_z]^T$.

The mathematical model used in this paper is proposed by Ateşoğlu, et al. in [19]. The dynamical equations for the two-axis gimbal system in the matrix form are provided in Eqn. 1. Given the driving torques, $T_a$ and $T_e$, the relative angular accelerations of the azimuth and elevation gimbals, $\ddot{\psi}_a$, $\ddot{\theta}_m$, can be calculated. Furthermore, the reaction forces ($\bar{F}_{a/m}^{(a)}$, $\bar{F}_{a/b}^{(a)}$) and the constraining components of reaction moments ($M_{amx}, M_{amz}, M_{abx}, M_{aby}$) acting on the gimbal revolute joints are calculated as well from the mathematical model.

$$\hat{F}_{12x2}\begin{bmatrix}\ddot{\psi}_a\\\ddot{\theta}_m\end{bmatrix}_{2x1} + \hat{R}_{12x10}\begin{bmatrix}\bar{F}_{a/m}^{(a)}\\\bar{F}_{a/b}^{(a)}\\M_{amx}\\M_{amz}\\M_{abx}\\M_{aby}\end{bmatrix}_{10x1} = \hat{D}_{12x1} + \hat{G}_{12x2}\begin{bmatrix}T_a\\T_e\end{bmatrix}_{2x1} \quad (1)$$

In addition, the commanded torques, $T_{ac}$, $T_{ec}$, can be calculated by inverting Eqn. 1, when the reference values of angular accelerations, $\ddot{\psi}_{ac}, \ddot{\theta}_{mc}$, angular velocities, $\dot{\psi}_{ac}, \dot{\theta}_{mc}$, and angular positions, $\psi_{ac}, \theta_{mc}$, of the azimuth and elevation gimbals are available. This is shown in the block diagram given in Fig. 2, and named as the *Computed Torque Model*.

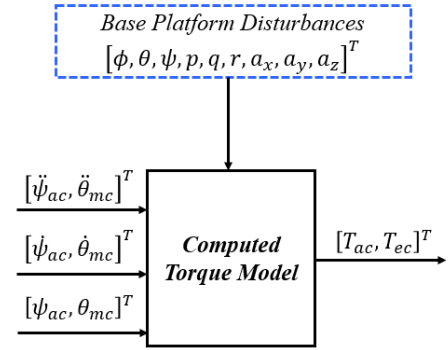

Fig. 2. Block diagram of the *Computed Torque Model*

## III. CONTROLLER DESIGN

### A. Classical ADRC with CTM

The classical ADRC is composed of three parts: *Transient Profile Generator (TG), Weighted State Error Feedback (WSEF)* and *Total Disturbance Estimation and Rejection* by using *ESO*. This controller is designed and constructed for distinct command channels seperately. Equations for a single channel are given below. Also, the block diagram representation of the ADRC structure for the elevation axis is given in Fig. 3 (ADRC structure for the azimuth axis is similar).

$$\dot{v}_1 = v_2 \qquad (2)$$
$$\dot{v}_2 = -rsign(v_1 - v + \frac{v_2|v_2|}{2r})$$

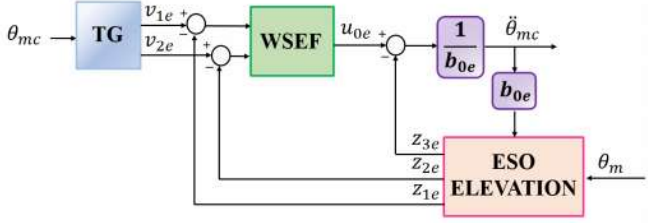

Fig. 3. TG, WSEF and ESO structure for the elevation axis

*Transient Profile Generator (TG):*
This part introduces the implementation of a high-quality nonlinear derivative operation without noise. Here, $v_1$ is the desired trajectory, $v_2$ is its derivative, $v$ is the reference input (desired position references, $\psi_{ac}$ and $\theta_{mc}$), $r$ is the learning rate that speeds up or slows down the transient profile and it needs to be tuned for perfect derivative action.

*Extended State Observer (ESO):*
Consider the state space respsentation of the nonlinear system expressed in Eqn. 3. An additional state variable $x_3$ is created to represent total disturbances, $F(t)$ in the system where $\dot{F}(t) = G(t)$, and $G(t)$ is unknown.

$$\begin{aligned}\dot{x}_1 &= x_2 \\ \dot{x}_2 &= x_3 + bu, \; x_3 = f(x_1, x_2, w(t), t) = F(t) \\ \dot{x}_3 &= G(t) \\ y &= x_1\end{aligned} \quad (3)$$

The state observer (ESO), is constructed with the set of equations given in Eqn. 4. $z_1$, $z_2$ and $z_3$ are the outputs of ESO. $z_1$ is the observed position, $z_2$ is the observed velocity and $z_3$ is the estimated disturbance. Explicit form of the nonlinear *fal(.)* function is given in Eqn. 5. It is used to eliminate the chattering problem highly possible to arise from the *sign()* function. The observer parameters needed to be tuned are $\beta_{01}, \beta_{02}, \beta_{03}, \alpha_1, \alpha_2, \delta$. Furthermore, a scaling parameter $b_0$, should also be tuned. In this study, the control input is $u$ ($\ddot{\psi}_{ac}, \ddot{\theta}_{mc}$) and plant output is $y$ ($\psi_a, \theta_m$). Also, following [10], $\alpha_1 = 0.5$ and $\alpha_2 = 0.25$ are used.

$$\begin{aligned}e &= z_1 - y, fe = fal(e, \alpha_1, \delta), fe_1 = fal(e, \alpha_2, \delta) \\ \dot{z}_1 &= z_2 - \beta_{01} e \\ \dot{z}_2 &= z_3 + b_0 u - \beta_{02} fe \\ \dot{z}_3 &= -\beta_{03} fe_1\end{aligned} \quad (4)$$

$$fal(e, \alpha, \delta) = \begin{cases} \dfrac{e}{\delta^{1-\alpha}}, & |e| \leq \delta \\ |e|^\alpha sign(e), & |e| > \delta \end{cases} \quad (5)$$

*Weighted State Error Feedback (WSEF):*
Here, WSEF acts like a simple PD (proportional derivative) type controller. Commands generated from Eqn. 2, ($v_1$ and $v_2$), are used to form the error signals ($e_1$ and $e_2$). Weighted combinations of error signals form the reference acceleration input ($u_o$) for the *CTM*. $k_1$ and $k_2$ are the parameters to be tuned. The block diagram in Fig. 4 shows the signal flow for $\bar{v}_1, \bar{v}_2, \bar{z}_1$ and $\bar{z}_2$, to form $\bar{e}_1$ and $\bar{e}_2$ (bar denotes concatenation of azimuth and elevation vectors). Note that $\bar{v}_1$ and $\bar{v}_2$ will be the position and velocity inputs for the *CTM*.

$$\begin{aligned}e_1 &= v_1 - z_1 \\ e_2 &= v_2 - z_2 \\ u_o &= k1 e_1 + k2 e_2 \\ u &= (u_0 - z_3)/b_0\end{aligned} \quad (6)$$

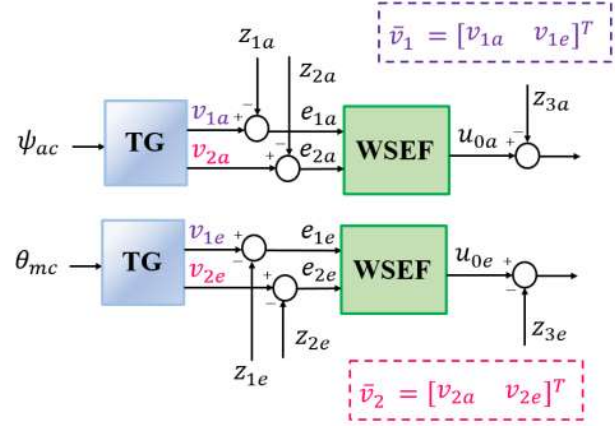

Fig. 4. TG and WSEF for azimuth and elevation gimbals

ADRC with *CTM* structure is given in Fig. 5 (plant block is composed of *Gimbal Dynamics* and the DC motor model). $u_{0a}$ and $u_{0e}$ are the outputs of WSEF blocks from azimuth and elevation axes, respectively. The CTM block is named as *ideal* (described with $\bar{D} = \bar{0}$). There is no (i) friction torque in revolute joints, (ii) disturbance torque due to electrical cables of motors, encoders, gyroscope, camera and cooling pipes, (iii) disturbance torque due to off-diagonal inertia terms, rotation axes and CoG (center of gravity) offsets. However, the plant is *non-ideal*, which means, $\bar{D} \neq \bar{0}$.

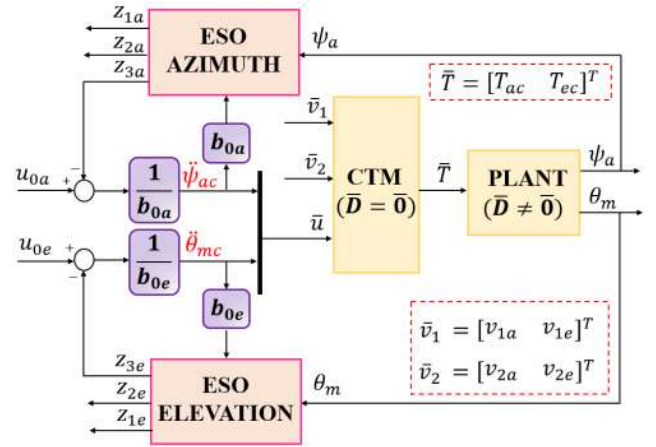

Fig. 5. ADRC with CTM structure

B. *NN Based ADRC Implementation*

As it is seen before, ADRC has certain parameters to be tuned. This paper proposes an intelligent controller, NN assisted ADRC structure which eases calibration and tuning of the legacy controller. It also unlocks the possible need for scheduling of the controller parameters which can be needed to cope with the time varying disturbances on the system and, or different reference signal levels. NN is employed to approximate the complex nature of possible disturbances which are present in the hardware set-up (Fig. 1) and enhance the tracking performance of the already constructed legacy ADRC structure. Addition of the proposed NN also speeds up the response and decreases tracking error on the position control loop. The performance of the proposed controller is verified by simulations presented in Section IV.

NN training is performed with the block diagram shown in Fig. 6. Please pay attention to the fact that, there are no disturbances in the CTM block used for training. The difference between $[\ddot{\psi}_{ac}\ \ddot{\theta}_{mc}]^T$ and $[\ddot{\psi}_{a}\ \ddot{\theta}_{m}]^T$ is accounted as the compensatory acceleration that needs to be fed on the control commands generated by the legacy controller.

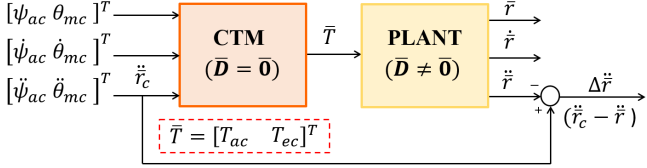

Fig. 6. NN training structure

The training data set of the neural network is $\{(\bar{r},\dot{\bar{r}},\ddot{\bar{r}}), \Delta\ddot{\bar{r}}, t \in [0, t_f]\}$, $\bar{r} = [\psi_a\ \theta_m]^T$, $\dot{\bar{r}} = [\dot{\psi}_a\ \dot{\theta}_m]^T$ and $\ddot{\bar{r}} = [\ddot{\psi}_a\ \ddot{\theta}_m]^T$. Output of the NN is the compensatory acceleration, $\Delta\ddot{\bar{r}} = [\Delta\ddot{\psi}_a\ \Delta\ddot{\theta}_m]^T$. The block diagram representation of the NN based ADRC controller is given in Fig. 7. $\Delta\ddot{\bar{r}}$ enables the plant to reach the reference position in a closed loop structure when it is added to $[\ddot{\psi}_{ac}\ \ddot{\theta}_{mc}]^T$. With the implementation of NN, $u = \left(\frac{u_0 - z_3}{b_0}\right) + \Delta r$.

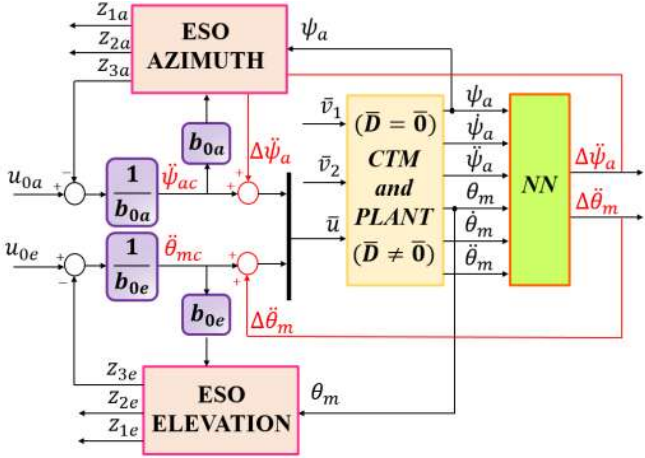

Fig. 7. NN based ADRC structure

### C. CTM Based ADRC Implementation

As an alternative, instead of using a NN supported ADRC controller, complementary torque signals can also be generated from a second *CTM block* in the controller architecture. By using the output of the plant as the input for the second *CTM block*, the differential torque between the first *CTM* and the second *CTM* can be calculated and added to the system. With this approach, instead of training a NN with the data obtained from *CTM* and plant, the compensatory torque signals can be directly supplied from the secondary *CTM block*. Support will be in the form of delta torque intsead of delta acceleration. The block diagram representation of the *CTM* assisted ADRC controller is given in Fig. 8. The total torque applied to the plant is the summation of the output of the first *CTM block* ($\bar{T}$) and the difference between the outputs of the first and second *CTM blocks* ($\Delta\bar{T}$).

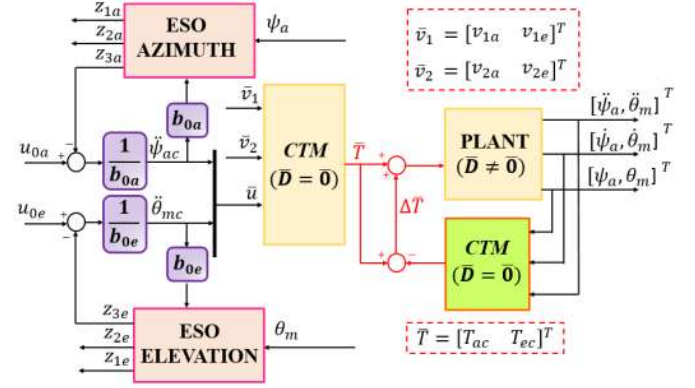

Fig. 8. CTM based ADRC structure

## IV. SIMULATIONS

In this section, we present the performance of the NN assisted ADRC formulation presented in Sec. III. Experiments of the system with NN-ADRC, CTM-ADRC hybrid controllers and conventional legacy ADRC controller are carried out by using MATLAB® and Simulink® R2020b. Equations of gimbal dynamics and CTM are implemented as MATLAB® function blocks. The sampling period of the simulated system is chosen to be 0.001 seconds and the numerical integration method is ODE4 (Runge-Kutta). Properties of the simulated two-axis gimbal system, calculated from Catia v5-3DX® are given in Tabs. I and II. $\bar{r}_{Ga/a}^{(a)}, \bar{r}_{Gm/m}^{(m)}, \bar{r}_{m/a}^{(a)}, \bar{r}_{a/b}^{(b)}$ in Tab. I, are the distance vectors between revolute joints and CoG points and revolute joints. Tab. II presents the mass, inertia and field-of-regard (FOR) limits of the azimuth and elevation gimbals ($J_{ii}$ are the diagonal inertia terms along directions $i = x, y, z$). Off-diagonal inertia terms calculated from 3D model are in $10^{-6}$ or $10^{-5}$ order (they are approximately twice smaller than primary diagonal inertia terms). Although they are small enough, the off-diagonal terms are included in the gimbal models (dynamical mass unbalance is included).

| *Distances (in mm)* | |
|---|---|
| $\bar{r}_{Ga/a}^{(a)} = [Ga_x\ Ga_y\ Ga_z]^T$ | $[0\ 0\ 57.5]^T$ |
| $\bar{r}_{Gm/m}^{(m)} = [Gm_x\ Gm_y\ Gm_z]^T$ | $[0\ -44.5\ 0]^T$ |
| $\bar{r}_{m/a}^{(a)} = [am_x\ am_y\ am_z]^T$ | $[0\ 44.5\ 57.5]^T$ |
| $\bar{r}_{a/b}^{(b)} = [ba_x\ ba_y\ ba_z]^T$ | $[31.625\ 0\ -57.5]^T$ |

TABLE I: Distances of the gimbal platform

|  | *Yaw Gimbal* | *Pitch Gimbal* |
|---|---|---|
| $m\ (kg)$ | 0.55 | 1.138 |
| $\hat{J}\ (kgm^2)$ | $J_{xx} = J_{zz} = 0.002$<br>$J_{yy} = 0.004$ | $J_{xx} = J_{zz} = 0.004$<br>$J_{yy} = 0.003$ |
| **FOR limits** | ±45° | ±20° |

TABLE II: Parameters of the gimbal platform

The feedforward neural network designed for the simulations has 2 hidden layers, each composed of 20 neurons. Layer outputs are calculated by the transfer function ***poslin (positive linear).*** The training method is chosen as ***Levenberg-Marquardt*** backpropagation. The disturbance torque in the system is given in Eqn. 7, $(x = a, m)$. The terms in the disturbance function are single and dual combinations of $\theta_m, \dot{\theta}_m, \psi_a, \dot{\psi}_a, \theta_m^2, \psi_a^2, \dot{\theta}_m^2, \dot{\psi}_a^2$. The state independent bias term ($k37_x$) is included as well. Coefficients are determined by curve fitting using data from the experimental set-up. The parameters $k1_x \dots k36_x$ are associated with (i), (ii), and the parameter $k37_x$ is associated with (iii), in Section III.A.

$$T_{dx} = k1_x \dot{\theta}_m + k2_x \theta_m + \cdots + k34_x(\dot{\psi}_a^2 \theta_m^2) + k35_x(\dot{\psi}_a^2 \dot{\theta}_m^2) + k36_x(\dot{\theta}_m^2 \theta_m^2) + k37_x \quad (7)$$

The generic base motion data history, $(p, q, r, a_x, a_y, a_z)$, is given in Fig. 9. It is obtained from a real-time flight of the carrier vehicle. For the following examples given below, NN is trained based on the data under amplitude and frequency varying sinusoidal excitations that travel almost all ranges in the FOR limits of the experimental set-up. The reference position travels from 5° to 25° for the azimuth axis and from 2° to 18° for the elevation axis, with an increasing frequency of 0.5 to 4 Hz. MATLAB® Neural Network Toolbox is used for training this network.

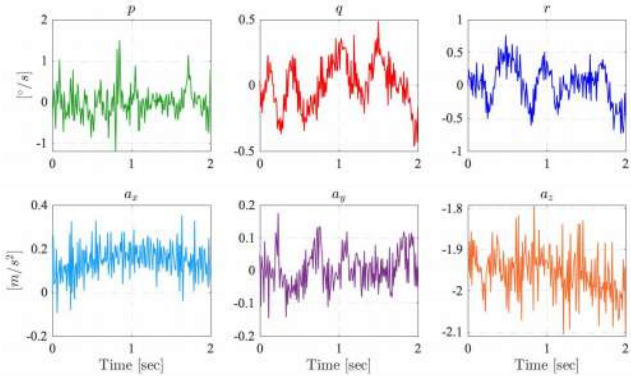

Fig. 9. Base platform disturbance data

In the first example, both azimuth and elevation gimbals are excited with the reference (*): 5°@ 1Hz sine wave. Fig. 10 gives the trajectory tracking response of the azimuth gimbal and Fig. 11 shows the tracking error. It has been observed that, without NN and CTM block, azimuth gimbal reaches to 4.55°. NN supports the ADRC controller and helps the gimbal to reach the reference position (5°) with a faster tracking rate. The mean track error (MTE) of ADRC is 0.09°. The MTE and percent improvement of hybrid controllers are given in Tab. III (% decrease indicates the MTE enhancement wrt. legacy ADRC).

|  | **NN-ADRC** | **CTM-ADRC** |
|---|---|---|
| **MTE** | 0.02° | 0.05° |
| **% decrease** | 76% | 37% |

TABLE III: MTE Comparison for Example 1

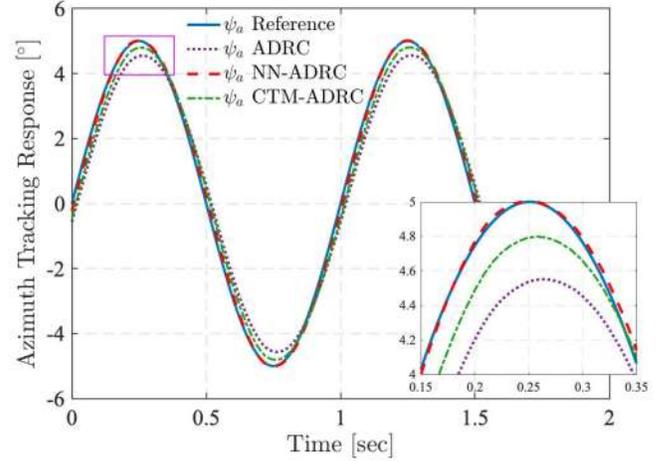

Fig. 10. Azimuth gimbal position for the reference (*)

As shown in Fig. 11, supplementary acceleration and torque provided by the NN and CTM, decreases tracking error. Performance of CTM-ADRC is in between ADRC and NN-ADRC.

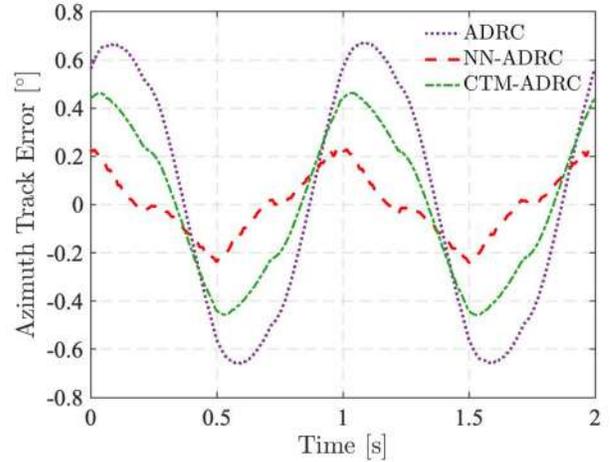

Fig. 11. Azimuth gimbal track error for the reference (*)

In the second example, azimuth and elevation gimbals are excited with the same reference (*). Disturbance torques of azimuth and elevation gimbals are increased respectively by 2 and 3 times wrt. the case in Example 1. Even though disturbance functions include model parameter uncertainties represented by the term $k37_x$, CoG points of the azimuth and elevation gimbals are also distorted further; $Ga_y = Gm_z = -0.01$, (i.e., they are different from zero). $J_{xy}, J_{yx}$ terms of the azimuth gimbal are $0.02\ kgm^2$ (this is a relatively large disturbance). In total, original model parameters extracted from the CAD model are highly distorted.

In this example, the NN trained with the previous example's data is used for the first three seconds. Then, it is switched to the NN based on the new disturbance data. Since ADRC is tuned wrt. old disturbance, control accuracy of the ADRC drops considerably (MTE increases from 0.09° to 0.16°). When NN-ADRC is used with the old neural network, it still has a better performance than ADRC (% decrease is 35.3%). % decrease of CTM-ADRC is slightly better than the old NN (40.8%). On the other hand, there is no chance to improve the performance of CTM controller further. However, we cannot talk about NN being optimal. With the adjustment of NN, performance of the NN-ADRC returns back to its old value. This suggests that a better strategy should be to train the NN online with the last available data. Real-time training of NN makes the system robust and insensitive to environmental/internal changes of disturbance torque. The MTE and % decrease of hybrid controllers after the adjustment of NN are given in Tab. IV. Figs. 12 and 13 show the tracking response and error of azimuth gimbal.

|            | NN-ADRC | CTM-ADRC |
|------------|---------|----------|
| **MTE**    | 0.023°  | 0.095°   |
| **% decrease** | 85.4% | 40.8%  |

TABLE IV: MTE Comparison for Example 2

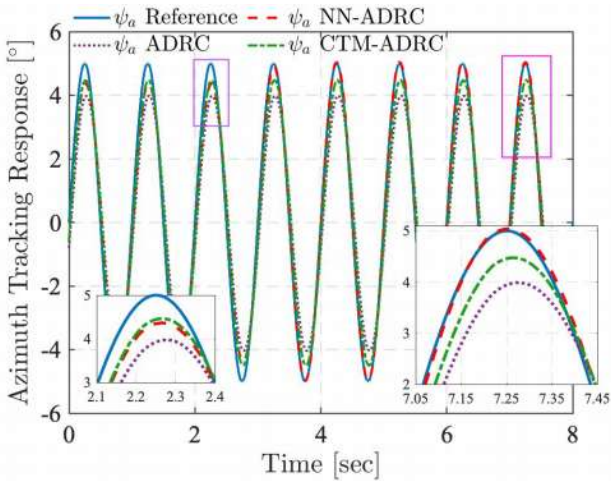

Fig. 12. Azimuth gimbal position for the reference (*): Example 2

## V. CONCLUSION AND FUTURE WORK

In this paper, a controller with a supplementary compensator (neural network) is designed to deal with the disturbance torque present in the real system and support the already constructed legacy ADRC controller. It performs MIMO based compensation for the cross-coupled gimbal model. The NN-ADRC control method reduces the time and effort required for tuning the parameters of ADRC controller. Also, CTM based ADRC structure is investigated as a possible alternative for NN-ADRC. Comparative results to check the performances of suggested controllers are provided by simulation studies and presented by figures and tables. Those results indicate that, NN-ADRC hybrid controller can decrease MTE up to 85.4% when compared to legacy ADRC (Tab. II). Moreover, it is superior when compared with a possible alternative, i.e., the CTM-ADRC, it can decrease MTE 4.5 times more than the CTM-ADRC controller (Tab. II).

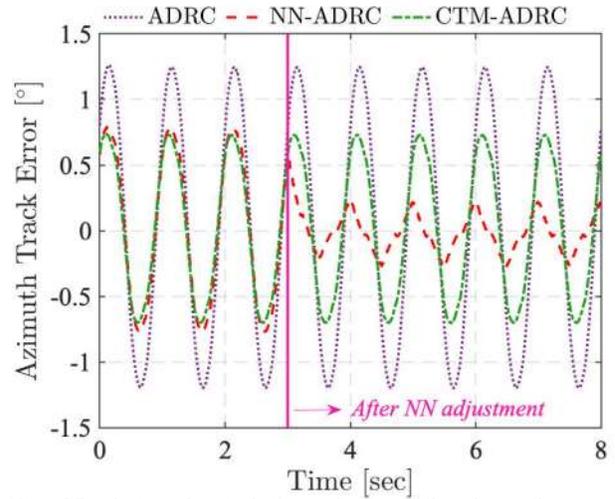

Fig. 13. Azimuth gimbal track error for the reference (*): Example 2

Real time adjustment of the weights of NN, gives the proposed algorithm strong robustness and environmental adaptability. We are planning to train the NN online and implement both online/offline NNs on the physical system using xPC Target®, as a future study.

ACKNOWLEDGMENT

This study is a joint work between Roketsan Missiles Inc. and Bilkent University, Department of Mechanical Engineering. Authors would also like to thank Mr. Anıl E. Derinöz for his technical support and discussions.